\documentclass[a4paper]{article}

\usepackage{INTERSPEECH2022}
\usepackage{cite}
\usepackage{subcaption}
\usepackage{multirow}
\usepackage{booktabs}
\usepackage{bm}
\usepackage{datatool}
\usepackage{enumitem}

\DTLsetseparator{,}
\DTLloaddb{tab_nat1_grp}{results/tab_nat1_grp.csv}
\DTLloaddb{tab_pt1_grp}{results/tab_pt1_grp.csv}
\DTLloaddb{tab_nat2_grp}{results/tab_nat2_grp.csv}
\DTLloaddb{tab_pt2_grp}{results/tab_pt2_grp.csv}

\newif\ifCommentsAuthors
\CommentsAuthorstrue
\usepackage{xcolor}
\ifCommentsAuthors
    \usepackage[colorinlistoftodos]{todonotes}
    \definecolor{myblue}{rgb}{0,0,.8}
    \newcommand{\commentjulz}[1]{\textcolor{myblue}{#1}}
    \definecolor{marccolor}{rgb}{0,0.5,0.5}
    \newcommand{\commentmarc}[1]{\textcolor{marccolor}{#1}}
    \definecolor{mcolor}{rgb}{0.5,0.2,0.1}
    \newcommand{\commentbenj}[1]{\textcolor{mcolor}{#1}}
    \definecolor{mygreen}{rgb}{.0,.8,.0}
    \newcommand{\commenthugo}[1]{\textcolor{mygreen}{#1}}
\else
    \definecolor{myred}{rgb}{.8,.0,.0}
    
    \newcommand{\commentmarc}[1]{}
    \newcommand{\commentjulz}[1]{}
    \newcommand{\commentbenj}[1]{}
    \newcommand{\commenthugo}[1]{}
\fi

\title{Daft-Exprt: Cross-Speaker Prosody Transfer on Any Text \\ for Expressive Speech Synthesis}
\name{Julian Zaïdi$^1$, Hugo Seuté$^1$, Benjamin van Niekerk$^{12}$, Marc-André Carbonneau$^1$}
\address{
  $^1$Ubisoft La Forge, Montreal, Canada\\
  $^2$Stellenbosch University, Stellenbosch, South Africa}
\email{\{julian.zaidi, hugo.seute, benjamin.van-niekerk, marc-andre.carbonneau2\}@ubisoft.com}

\begin{document}
\begin{filecontents*}{results/tab_nat1_grp.csv}
model,mean,+/-
audio_ref,4.42,0.05
audio_ref_HiFi,3.85,0.07
low_anchor_HiFi,2.25,0.08
model0_HiFi,3.28,0.08
model3_HiFi,3.47,0.08
model4_HiFi,2.86,0.08
model5_HiFi,3.36,0.08
\end{filecontents*}

\begin{filecontents*}{results/tab_pt1_grp.csv}
model,mean,+/-
low_anchor_HiFi,12.7,1.1
model0_HiFi,58.0,2.0
model3_HiFi,36.2,1.9
model4_HiFi,40.9,2.0
model5_HiFi,37.9,1.8
\end{filecontents*}

\begin{filecontents*}{results/tab_nat2_grp.csv}
model,mean,+/-
audio_ref,4.41,0.04
low_anchor_HiFi,2.37,0.08
model0_HiFi,3.24,0.08
model1_HiFi,2.88,0.08
model2_HiFi,3.72,0.07
\end{filecontents*}

\begin{filecontents*}{results/tab_pt2_grp.csv}
model,mean,+/-
low_anchor_HiFi,17.3,1.2
model0_HiFi,55.0,1.8
model1_HiFi,61.1,1.8
model2_HiFi,35.2,1.7
\end{filecontents*}

\maketitle

\begin{abstract}
This paper presents Daft-Exprt, a multi-speaker acoustic model advancing the state-of-the-art for cross-speaker prosody transfer on any text. This is one of the most challenging, and rarely directly addressed, task in speech synthesis, especially for highly expressive data. Daft-Exprt uses FiLM conditioning layers to strategically inject different prosodic information in all parts of the architecture. The model explicitly encodes traditional low-level prosody features such as pitch, loudness and duration, but also higher level prosodic information that helps generating convincing voices in highly expressive styles. Speaker identity and prosodic information are disentangled through an adversarial training strategy that enables accurate prosody transfer across speakers. Experimental results show that Daft-Exprt significantly outperforms strong baselines on inter-text cross-speaker prosody transfer tasks, while yielding naturalness comparable to state-of-the-art expressive models. Moreover, results indicate that the model discards speaker identity information from the prosody representation, and consistently generate speech with the desired voice. We publicly release our code\footnote{https://github.com/ubisoft/ubisoft-laforge-daft-exprt} and provide speech samples from our experiments\footnote{https://ubisoft-laforge.github.io/speech/daft-exprt}.

\end{abstract}
\noindent\textbf{Index Terms}: speech synthesis, prosody transfer

\section{Introduction}
\label{sec:intro}

The first wave of neural text-to-speech systems showed the ability to synthesize natural-sounding speech \cite{Wang2017, Shen2017, Li2019}. However, these systems generate speech from text alone, offering no control over prosody (i.e. all speech information that is not contained in text, speaker identity and channel effects \cite{Skerry-Ryan2018}). Harnessing prosody is instrumental for providing expressive synthetic voices in entertainment applications such as movies and video games, or to increase user engagement in human-machine interactions \cite{Aarestrup2015}. As a result, many recent efforts are devoted towards accurate and expressive prosody transfer.

Prosody transfer uses a reference utterance to condition the generation of a speech line. The speaker of the reference line can be the same as the targeted speaker or different. Cross-speaker prosody transfer is much more challenging because the system needs to represent prosody and speaker identity information independently. Similarly, reference and target texts can be the same or different. Same-text cross-speaker methods capture time-dependent attributes from the reference utterance and offer fine-grained prosody transfer \cite{Lee2018, Karlapati2020, Valle2020a}. However, being constrained to the same, or very similar, text limits the usefulness of these approaches for specific applications. For inter-text cross-speaker transfer, some methods capture time-independent and high-level attributes from the reference utterance. For instance, in \cite{Li2021}, a reference encoder extracts prosodic information to learn style discriminative embeddings, while \cite{Bian2019} and \cite{Whitehill2019, An2021} use a multi-reference approach to combine audio of different styles in an intercross and adversarial training scheme. A drawback of these methods is that they require data sets where styles are pre-defined (e.g. angry, newscaster). It is unclear how these methods deal with large variations across the same style label and how well they scale with the number of style labels, which is required to cover the full range of human prosody. 

\begin{figure}[t]
    \centering
    \includegraphics[width=0.95\linewidth]{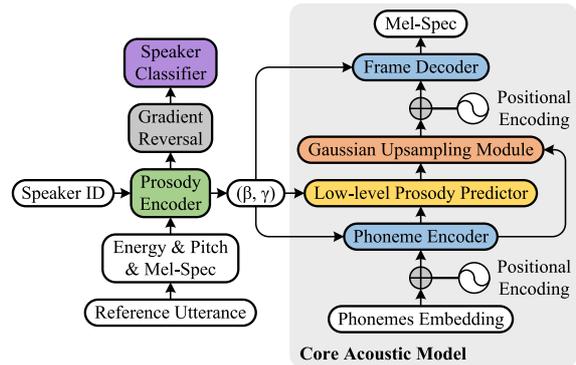}
    \caption{Daft-Exprt architecture \label{fig:general_architecture}}
\end{figure}

\begin{figure*}[h]
  \begin{subfigure}[t]{.2\linewidth}
    \centering
    \includegraphics[width=0.83\linewidth]{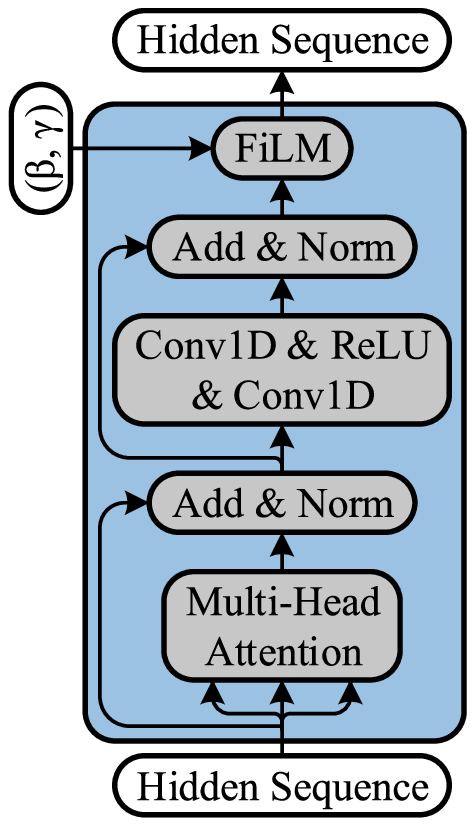}
    \caption{FFT block \label{fig:FFT_block}}
  \end{subfigure}
  \begin{subfigure}[t]{.2\linewidth}
    \centering
    \includegraphics[width=0.85\linewidth]{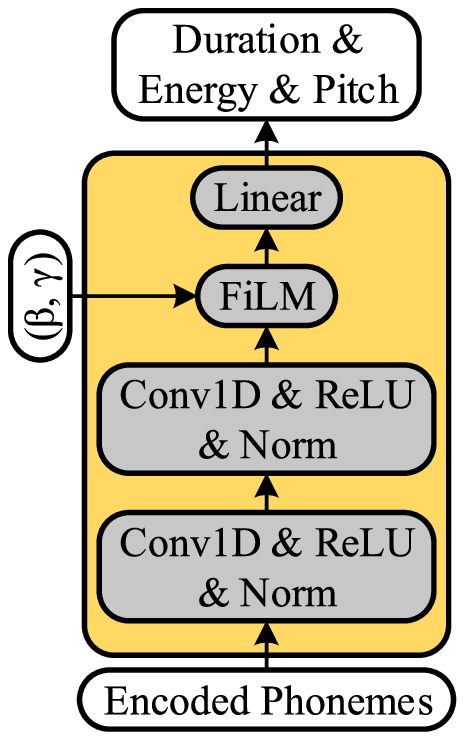}
    \caption{Low-level Prosody  \label{fig:LPP}}
  \end{subfigure}
  \begin{subfigure}[t]{.25\linewidth}
    \centering
    \includegraphics[width=0.98\linewidth]{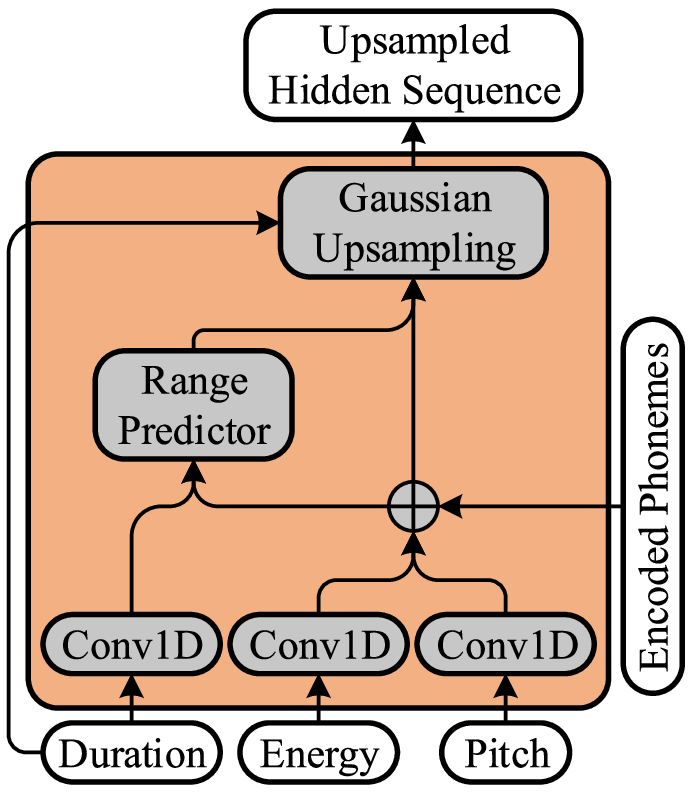}
    \caption{Gaussian Upsampling \label{fig:GU}}
  \end{subfigure}
  \begin{subfigure}[t]{.3\linewidth}
    \centering
    \includegraphics[width=0.95\linewidth]{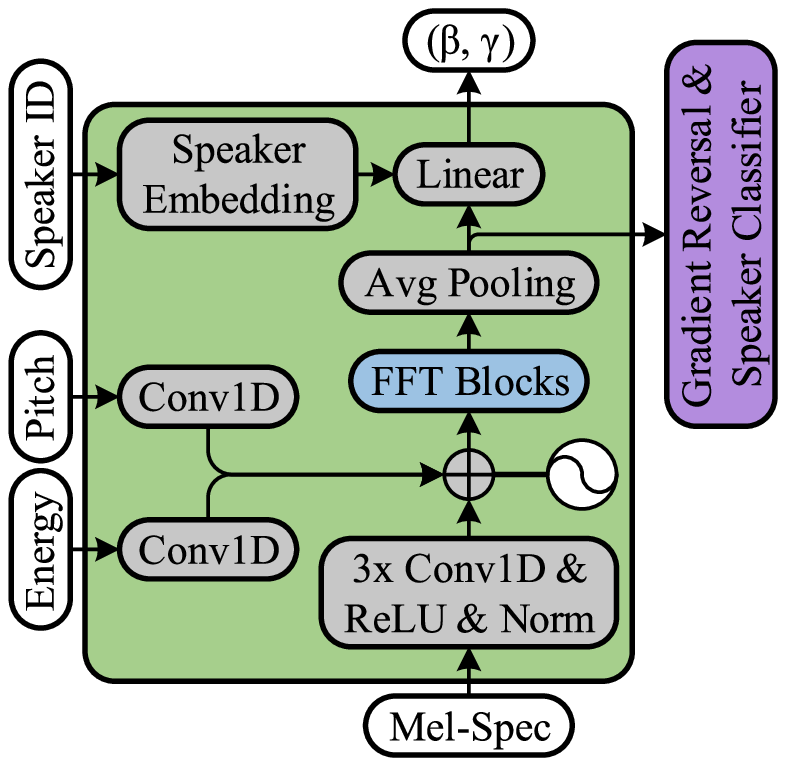}
    \caption{Prosody Encoder \label{fig:PE}}
  \end{subfigure}
  \caption{Daft-Exprt modules \label{fig:modules}}
\end{figure*}

Several methods can learn to transfer prosody without labeled style data sets. One popular approach is to capture high-level prosodic information in a latent space. For instance, \cite{Wang2018} use a mixture of learned style prototypes to encode the reference utterance, while \cite{Zhang2019a, Hsu2018, Battenberg2019} and \cite{Valle2020b} use variational and flow-based approaches. These architectures perform high-level prosody transfer, but they do not explicitly model low-level prosodic features, as done in \cite{Ren2020, Ancucki2020, Shen2020}, which hinders their accuracy when reproducing timing-sensitive events (e.g. pitch raising at the end of a question). In theory, these methods allow for inter-text cross-speaker transfer, but formal empirical studies still lack in the literature. The model in \cite{Lee2021} summarizes prosody as a set of low-level features, namely pitch, loudness, duration and noise. However, prosody is much more than these features, and the missing high-level prosodic information limits the inter-text cross-speaker transfer capabilities of the model to neutral-ish styles only. For instance, from these low-level features, it is difficult to tell if a line should be whispered, shouted, said in a raspy voice, etc. This motivated us to propose a model that explicitly condition and predicts low-level prosodic features to encourage the capture of timing-sensitive information, in conjunction with traditional higher-level conditioning to transfer time-independent prosodic information related to voice effort, harshness, breathiness, etc.

One of the main challenges for prosody transfer across speakers is to disentangle prosodic and speaker identity information. Learned latent prosodic representations tend to encode speaker identity with prosody \cite{Skerry-Ryan2018, Battenberg2019}, especially when prosodic styles are not uniformly distributed across speakers in the data. This results in the so-called speaker leakage problem \cite{Karlapati2020}, where prosody is accurately transferred, but the synthesized speech exhibits unwanted attributes from the reference speaker. Adversarial training has been used before in speech applications to discard unwanted information from a latent representation. For instance, in \cite{Zhang2019b}, an adversarial speaker classifier encourages disentanglement between speaker identity and linguistic content. Similarly, \cite{Shang2021} disentangle speaker identity and accent, which in our context can be interpreted as a prosodic style. Inspired by these efforts, we propose to use adversarial training to learn speaker-invariant prosodic representations.

Model conditioning strategies are still being actively researched. Conditioning acoustic models with prosody is generally achieved through concatenation \cite{Lee2018, Karlapati2020, Valle2020a, Li2021, Bian2019, Whitehill2019, An2021, Wang2018, Zhang2019a, Hsu2018, Battenberg2019, Valle2020b}. While this is an effective method, feature-wise affine transformation on normalization layers have improved the state-of-the-art in several domains (e.g. image stylization \cite{Dumoulin2017, Huang2017} or speech \cite{Kim2017, Chen2021a, Chen2021b, Min2021}). This technique, also called conditional normalization, was later generalized by FiLM layers \cite{Perez2018} as a general-purpose conditioning method that is decoupled from normalization operations. We propose to use FiLM conditioning for their expressive power, and to independently condition each part of the acoustic model.

This paper presents Daft-Exprt, a multi-speaker acoustic model capable of accurate prosody transfer across speakers. Daft-Exprt stands for \textbf{D}eep \textbf{af}fine \textbf{t}ransformations for \textbf{Ex}pressive \textbf{pr}osody \textbf{t}ransfer. We target a difficult and rarely addressed problem setting, where we transfer highly expressive prosodic styles on any text from seen and unseen reference speakers.  Using FiLM layers, Daft-Exprt can transfer detailed prosodic information from a reference utterance by conditioning independently all parts of the acoustic model. This enables the model to encode low-level prosodic information for the pitch, duration and loudness predictors, as well as providing higher level prosodic information to the frame decoder. In addition, Daft-Exprt uses adversarial training to encourage disentanglement between prosody and speaker identity, improving target speaker fidelity for cross-speaker prosody transfer. We compare Daft-Exprt against 3 strong baselines on a highly expressive multi-speaker data set, and demonstrate accurate prosody transfer in challenging inter-text cross-speaker experiments. We provide samples showcasing expressive prosody transfer, using seen and unseen reference speakers, even on target speakers trained only with neutral data.

\section{Proposed Model}
\label{sec:proposed_model}

Daft-Exprt consists of a prosody encoder and a core acoustic model, as illustrated in Figure \ref{fig:general_architecture}. The core acoustic model generates spectrogram frames from a phoneme sequence. The prosody encoder applies feature-wise affine transformations to the core acoustic model based on a reference utterance.

\subsection{Core Acoustic Model}
\label{ssec:core_acoustic_model}

The core acoustic model builds on FastSpeech 2 \cite{Ren2020}, a non-autoregressive transformer-based model, and comprises four parts: a phoneme encoder, a low-level prosody predictor, a Gaussian upsampling module and a frame decoder.

The \textbf{phoneme encoder} extracts a sequential latent representation from an embedded phoneme sequence using Feed-Forward Transformer (FFT) blocks. A FFT block implements a multi-head self-attention and two 1D-convolution layers, each followed by a residual connection and layer normalization, as depicted in Figure \ref{fig:FFT_block}.

Next, the \textbf{low-level prosody predictor} estimates a duration, energy, and pitch scalar value for each phoneme. The architecture, shown in Figure \ref{fig:LPP}, is similar to the variance adaptor of \cite{Ren2020} but without the length regulator. All prosodic features are predicted using shared parameters. Following \cite{Ancucki2020}, we predict prosodic features at the phoneme level because control is more intuitive than at the frame level.

Daft-Exprt uses the \textbf{Gaussian upsampling module} in Figure \ref{fig:GU} to address the length mismatch between phoneme and spectrogram sequences. Gaussian upsampling \cite{Shen2020} replaces the length regulator of FastSpeech 2 to improve naturalness. First, three 1D-convolution layers project duration, energy and pitch to match the dimension of the phoneme representation. A range predictor then sums the projections and encoded phonemes before predicting a sequence of ranges using a linear layer with SoftPlus activation. Finally, a Gaussian upsampling layer adds the energy and pitch projections to the encoded phonemes and upsamples the sequence using integer frame durations as means and ranges as standard deviations.

As a last step, the \textbf{frame decoder} predicts spectrogram frames from the upsampled hidden sequence by applying several FFT blocks followed by a linear layer.

\subsection{Prosody Encoder}
\label{ssec:prosody_encoder}

The core acoustic model is conditioned in three places: the phoneme encoder, the low-level prosody predictor and the frame decoder. The \textbf{prosody encoder}, illustrated in Figure~\ref{fig:PE}, applies affine transformations to the intermediate features of specific layers. These transformations, illustrated as FiLM layers \cite{Perez2018} in Figure \ref{fig:FFT_block} and \ref{fig:LPP}, are conditioned on prosodic information extracted from a reference utterance in the form of energy, pitch and spectrogram sequences.

First, the spectrogram is encoded through three 1D-convolution layers, each followed by ReLU activation and layer normalization. Two 1D-convolution layers also project pitch and energy. The hidden sequences are then summed and fed to FFT blocks to produce a sequential intermediate representation. A latent \textbf{prosody vector}, summarizing the prosody of the whole utterance, is obtained by averaging over the sequence, and a \textbf{speaker embedding} is added to the prosody vector to specify the identity of the target voice. Finally, a linear layer predicts a $\beta$ and a $\gamma$ value for each conditioned feature in the core acoustic model. The $\gamma$ and $\beta$ values correspond to the scaling and bias performed by the FiLM layer. As in \cite{Oreshkin2018}, for regularization, each FiLM layer has a scaling value $s_{\gamma}$ and $s_{\beta}$ applied to all its $\gamma$ and $\beta$. These scaling values are not predicted, but learned during training.

\subsection{Training Loss}
\label{ssec:training_loss}

We use four loss terms for training: two regularization terms, $\mathcal{L}_{r}$ and $\mathcal{L}_{f}$, alongside a prediction error $\mathcal{L}_{e}$ and an adversarial loss $\mathcal{L}_{a}$. The relative importance of all terms is adjusted through $\lambda_f$, $\lambda_a$ and $\lambda_r$:
\begin{equation}
\mathcal{L} = \mathcal{L}_{e} + \lambda_{f}\mathcal{L}_{f} - \lambda_{a}\mathcal{L}_{a} + \lambda_r\mathcal{L}_{r}
    \label{eqn:loss}
\end{equation}

The prediction error $\mathcal{L}_{e}$ is the mean squared error for the mel-spectrogram, duration, energy and pitch predictions. We also add mean absolute error on mel-spectrogram predictions.

As in \cite{Oreshkin2018}, we observe that regularization ensures better generalization with FiLM conditioning. We follow \cite{Oreshkin2018} and use $\mathcal{L}_{f}$ to penalize the $\ell^2$-norm of the scaling parameters $s_{\gamma}$ and $s_{\beta}$. We also further regularize using weight decay $\mathcal{L}_{r}$.

During training, reference and target speakers are the same, which causes the prosody encoder to capture speaker identity. We encourage disentanglement between prosody and speaker identity using an adversarial loss term. $\mathcal{L}_{a}$ is the cross-entropy loss of a classifier trained to predict the speaker identity from the prosody vector, which we backpropagate through a gradient reversal layer \cite{Ganin2015}. In other words, the classifier penalizes the prosody encoder if the prosody vector contains information about the speaker from the reference utterance. The classifier consists of three linear layers with ReLU activations.

\section{Experiments and results}
\label{sec:experiments_and_results}

\subsection{Experimental Setup}
\label{ssec:experimental_setup}

The same training data is used in all experiments. We combine the LJ Speech data set \cite{LJSpeech17} with a 12 speaker internal data set containing 47 hours of recordings. We capture highly expressive performances in many styles from 7 of the speakers. The remaining speakers recorded only neutral lines.

We extract 80 bin mel-spectrograms from the recordings using a $\sim$46ms Hann window and a hop size of $\sim$12ms. We extract phoneme durations using MFA\footnote{http://github.com/MontrealCorpusTools/Montreal-Forced-Aligner}, estimate log-pitch with REAPER\footnote{https://github.com/google/REAPER}, and use the $\ell^2$-norm of the spectrogram frames to measure energy. We average energy and log-pitch per phoneme and standardize them per speaker.

The core acoustic model is configured as in \cite{Ren2020}. However, we use a kernel size of 3 for all convolution layers. We decrease the hidden dimension from 256 to 128 for the phoneme embedding, encoder and frame decoder for improved generalization.

We build the prosody encoder with a kernel size of 3 for all convolutions, and encode mel-spectrograms using 1024 channels. The prosody encoder maps inputs to a 128 dimension hidden representation, which is passed through 4 FFT blocks with 8 attention heads. Finally, we learn a 128 dimension speaker embedding. The speaker classifier used for adversarial training is also made of 128-dimensional layers.

We train Daft-Exprt with a batch size of 48 and use the Adam optimizer, configured as in \cite{Ren2020}. We linearly increase the learning rate from 1e-4 to 1e-3 during the first 10k steps before decaying. For the loss, we set $\lambda_{f}$ to 1e-3, $\lambda_{r}$ to 1e-6, and linearly increase $\lambda_{a}$ from 0 to 1e-2 during the first 10k steps. Finally, we use a dropout rate of 0.1 for all self-attention and convolution layers.

We compare Daft-Exprt against 3 strong prosody transfer baseline models: GST-Tacotron \cite{Wang2018}, VAE-Tacotron \cite{Zhang2019a} and Flowtron \cite{Valle2020b}. We add multi-speaker support to GST-Tacotron and VAE-Tacotron, as done in \cite{Valle2020b}. For Flowtron, we fine-tune a model pre-trained on LibriTTS \cite{Zen2019} that has been made available by the authors. All models are trained for 300 epochs on our data. Once training is finished, we fine-tune a pre-trained 22kHz universal HiFi-GAN vocoder\footnote{https://github.com/jik876/hifi-gan}\cite{Kong2020} for each acoustic model during 1M iterations.

\subsection{Evaluation Setup}
\label{ssec:evaluation_setup}

Our evaluations focus on expressive inter-text cross-speaker prosody transfer. We compare the models conditioning on 30 utterances from speakers seen during training and 30 utterances from speakers unseen during training. All the reference utterances used for evaluation are highly expressive and excluded from training data. For unseen speakers, we draw expressive samples from the 2013 Blizzard Challenge \cite{Blizzard2013} and RAVDESS \cite{Livingstone2018} data sets. For each reference utterance, we randomly select a text line and a target speaker that differ from the reference. We generate using all models and submit the samples to raters.

MUSHRA-like tests \cite{ITU2015} evaluate prosody transfer, while mean opinion scores (MOS) evaluate naturalness. As a low anchor for MUSHRA, we condition Daft-Exprt on a neutral reference utterance, randomize phoneme durations and flatten pitch curve. This yields a robotic sounding voice with an odd pacing. We discard raters giving high scores to low quality anchors and low naturalness scores to lines recorded by professional actors. We ensured that we maintained a minimum of 8 evaluations per audio after rejection. Raters are self-reported native English speakers. We report mean scores and 95\% confidence intervals.

Additionally, we compute Pearson correlation coefficient (PCC) between pitch curves of generated examples and the reference line. The lengths of the pitch curves differ since texts differ. To align both curves, we discard unvoiced segments and re-sample to get the same length. This allows us to measure the coarse similarity between both lines, independently of the fundamental frequency value. An accurate prosody transfer should translate into similar pitch fluctuation patterns between the reference and synthesized audio.

\subsection{Prosody Transfer}
\label{ssec:prosody_transfer}

Table \ref{tab:table1} reports MUSHRA scores obtained when comparing models on inter-text cross-speaker prosody transfer. For a fair evaluation between models, we explicitly ask raters to ignore speaker identity, audio quality or pronunciation mistakes, and to focus only on the prosody similarity between a reference utterance and synthesized speech samples.

\begin{table}[h]
\caption{MUSHRA, MOS and pitch ($F_0$) mean PCC for inter-text cross-speaker prosody transfer}
\label{tab:table1}
\centering
\begin{tabular}{l c c c}
\toprule
Method & MUSHRA & MOS & $F_{0}$ PCC  \\
\midrule
GST-Tacotron & $\DTLfetch{tab_pt1_grp}{model}{model4_HiFi}{mean} \pm \DTLfetch{tab_pt1_grp}{model}{model4_HiFi}{+/-}$ & $\DTLfetch{tab_nat1_grp}{model}{model4_HiFi}{mean} \pm \DTLfetch{tab_nat1_grp}{model}{model4_HiFi}{+/-}$ & 0.10 \\
VAE-Tacotron & $\DTLfetch{tab_pt1_grp}{model}{model5_HiFi}{mean} \pm \DTLfetch{tab_pt1_grp}{model}{model5_HiFi}{+/-}$ & $\DTLfetch{tab_nat1_grp}{model}{model5_HiFi}{mean} \pm \DTLfetch{tab_nat1_grp}{model}{model5_HiFi}{+/-}$ & 0.14 \\
Flowtron & $\DTLfetch{tab_pt1_grp}{model}{model3_HiFi}{mean} \pm \DTLfetch{tab_pt1_grp}{model}{model3_HiFi}{+/-}$ & $\DTLfetch{tab_nat1_grp}{model}{model3_HiFi}{mean} \pm \DTLfetch{tab_nat1_grp}{model}{model3_HiFi}{+/-}$ & 0.07 \\
Daft-Exprt & $\DTLfetch{tab_pt1_grp}{model}{model0_HiFi}{mean} \pm \DTLfetch{tab_pt1_grp}{model}{model0_HiFi}{+/-}$ & $\DTLfetch{tab_nat1_grp}{model}{model0_HiFi}{mean} \pm \DTLfetch{tab_nat1_grp}{model}{model0_HiFi}{+/-}$ & 0.43 \\
\midrule
GT + HiFi-GAN & - & $\DTLfetch{tab_nat1_grp}{model}{audio_ref_HiFi}{mean} \pm \DTLfetch{tab_nat1_grp}{model}{audio_ref_HiFi}{+/-}$ & - \\
Ground truth (GT) & - & $\DTLfetch{tab_nat1_grp}{model}{audio_ref}{mean} \pm \DTLfetch{tab_nat1_grp}{model}{audio_ref}{+/-}$ & - \\
\bottomrule
\end{tabular}
\end{table}

Results indicate a statistically significant improvement on prosody transfer when comparing Daft-Exprt to all baseline models. The large margin between scores is explained by the ability of Daft-Exprt to accurately transfer expressive prosodies, even to target speakers for which the data set contains only neutral utterances. Our demo page showcases some of these examples. We hypothesize that conditioning independently all parts of the core acoustic model with affine transformations translates into semantically meaningful modulations. Particularly, allowing a separate conditioning for the explicit prediction of low-level prosodic features encourages the model to better capture timing-sensitive information. For instance, while a rough estimate, the $F_{0}$ average PCC indicates that Daft-Exprt is better at reproducing reference pitch curves, which is important in the perception of prosody transfer accuracy.

\subsection{Naturalness}
\label{ssec:naturalness}

Naturalness MOS for all models and the ground truth recordings are reported in Table \ref{tab:table1}. Ground truth utterances are all expressive and are the ones used as references for prosody transfer in section \ref{ssec:prosody_transfer}. We also measure the impact of the vocoder on audio quality by evaluating re-synthesized ground truth spectrograms. Results show that the vocoder significantly degrades speech quality. As observed in \cite{Lorenzo2019}, vocoders produce more artifacts on highly expressive speech than on neutral voices.

VAE-Tacotron, Flowtron and Daft-Exprt get comparable results. However, advantages and sources of imperfection differ across models. The baseline models are auto-regressive, which has been shown to improve naturalness \cite{Watts2019}. On the other hand, they are prone to attention failure problems causing slurring, skipping or repetition of words. Attention failures are more frequent when generating in strong speaking styles than in neutral style. This explains why GST-Tacotron, rated as the best baseline for prosody transfer, was particularly subject to attention failures, and obtained a lower naturalness MOS. Finally, as stated earlier, vocoding expressive lines is more difficult than neutral and causes more artifacts. This means that models which transfer expressive prosodies more accurately, like Daft-Exprt and GST-Tacotron, are disadvantaged at the vocoding stage. We further discuss the impact of expressive transfer on synthesis naturalness in the next section.

\subsection{Prosody and Speaker Identity Disentanglement}
\label{ssec:prosody_and_speaker_identity_disentanglement}

As explained in Section \ref{ssec:training_loss}, the adversarial loss term encourages disentanglement between prosody and speaker identity. Here, we study the impact of this loss term by varying its importance weight $\lambda_{a}$ in equation \ref{eqn:loss}. We conduct MUSHRA tests to evaluate prosody transfer quality, as explained in section \ref{ssec:prosody_transfer}. Similar to \cite{Skerry-Ryan2018}, we also report the accuracy of a classifier predicting the identity of the target speaker from the generated spectrogram. The classifier consists of the prosody encoder in Section~\ref{ssec:prosody_encoder}, ignoring energy and pitch sequences, combined with the speaker classifier described in Section \ref{ssec:training_loss}. We train the classifier on ground truth mel-spectrograms until reaching a 100\% test set accuracy. We evaluate the accuracy for each model on 15k inter-text cross-speaker generated spectrograms.

\begin{table}[h]
\caption{MUSHRA, MOS and target speaker classification accuracy for Daft-Exprt trained with different adversarial weights}
\label{tab:table2}
\centering
\begin{tabular}{l c c c}
\toprule
  &  MUSHRA & MOS &  Accuracy  \\
\midrule
$\lambda_{a} = 0$  &  $\DTLfetch{tab_pt2_grp}{model}{model1_HiFi}{mean} \pm \DTLfetch{tab_pt2_grp}{model}{model1_HiFi}{+/-}$  & $\DTLfetch{tab_nat2_grp}{model}{model1_HiFi}{mean} \pm \DTLfetch{tab_nat2_grp}{model}{model1_HiFi}{+/-}$ & 40.5\%  \\
$\lambda_{a} = 0.01$  &  $\DTLfetch{tab_pt2_grp}{model}{model0_HiFi}{mean} \pm \DTLfetch{tab_pt2_grp}{model}{model0_HiFi}{+/-}$  & $\DTLfetch{tab_nat2_grp}{model}{model0_HiFi}{mean} \pm \DTLfetch{tab_nat2_grp}{model}{model0_HiFi}{+/-}$ & 97.6\%  \\
$\lambda_{a} = 1$  &  $\DTLfetch{tab_pt2_grp}{model}{model2_HiFi}{mean} \pm \DTLfetch{tab_pt2_grp}{model}{model2_HiFi}{+/-}$  & $\DTLfetch{tab_nat2_grp}{model}{model2_HiFi}{mean} \pm \DTLfetch{tab_nat2_grp}{model}{model2_HiFi}{+/-}$ & 100.0\%  \\
\bottomrule
\end{tabular}
\end{table}

Results in Table \ref{tab:table2} show how adversarial training impacts results. When $\lambda_{a}=0$, the prosody encoder is free to capture reference speaker identity information. The higher MUSHRA score indicates that prosody is accurately transferred. However, the lower classifier accuracy means that the model fails to generate with the desired speaker voice. As $\lambda_{a}$ increases, disentanglement is encouraged at the expense of accurate prosody transfer. We empirically searched for the best trade-off between prosody transfer and disentanglement and found $\lambda_{a} = 0.01$ yields best results. 

We also report naturalness MOS for each $\lambda_{a}$ configuration. Results concord with the findings in Section \ref{ssec:naturalness}. Generating expressive lines is more difficult for the acoustic model and the vocoder, and thus results in a lower naturalness. We observe an inverse relationship between prosody transfer accuracy and naturalness. When disentanglement is strongly enforced ($\lambda_{a} = 1$), the model ignores the reference prosody and generates lines that are closer to a neutral style, which are easier to predict and vocode, hence higher MOS (see example page). The adversarial loss encourages the prosody encoder to discard all information that could lead to speaker identification. Unfortunately, some prosodic information (e.g. speech rate) relates, to some extent, to speaker identity. Moreover, the data distribution of styles for each speaker can also be leveraged for speaker identification. This might lead the model to discard relevant prosodic information and to generate in the most common style, neutral.

\section{Conclusion}

This paper introduced Daft-Exprt, a multi-speaker acoustic model combining FiLM conditioning layers and adversarial training for accurate and expressive inter-text cross-speaker prosody transfer. Our experiments showed that Daft-Exprt advances the state-of-the-art on this difficult task, while delivering high quality speech. We also measured the effectiveness of adversarial training on disentanglement between speaker identity and prosody. As future work, we intend to replace our speaker encoding scheme to synthesize with any speaker, and increase spectrogram prediction accuracy for better sound quality.

\bibliographystyle{IEEEtran}

\bibliography{references}

\end{document}